\documentclass[letterpaper,twocolumn,10pt]{article}
\usepackage{usenix,epsfig,endnotes}
\usepackage{times}
\usepackage{epsfig}
\usepackage{graphicx}
\usepackage{amsmath}
\usepackage{amssymb}
\usepackage{multirow}
\usepackage{url}
\usepackage{pslatex}
\usepackage{graphics}
\usepackage{subfigure}
\usepackage{mathptmx}
\usepackage{verbatim}
\usepackage{textcomp}

\usepackage[pdftex,colorlinks=false,pdfborder={0 0 0},plainpages=true,pdfpagelabels=true]{hyperref}
\hypersetup
{
    pdfsubject={Scaling Turbo Boost to a 1000 cores},
    pdftitle={Scaling Turbo Boost to a 1000 cores},
}

\graphicspath{{figures/}}

\newcommand{\figref}[1]{Figure \ref{#1}}
\newcommand{\secref}[1]{Section \ref{#1}}
\newcommand{\tabref}[1]{Table \ref{#1}}

\begin{document}

\date{}

\title{ \Large  \bf Scaling Turbo Boost to a 1000 cores}
\author{
{\rm Ananth Narayan S.}\\
\texttt{ans6@sfu.ca}\\
Simon Fraser University\\
Canada
\and
{\rm Somsubhra Sharangi}\\
\texttt{ssa121@sfu.ca}\\
Simon Fraser University\\
Canada
\and
{\rm Alexandra Fedorova}\\
\texttt{fedorova@cs.sfu.ca}\\
Simon Fraser University\\
Canada
}

\maketitle

\subsection*{Abstract}
The Intel\textregistered~Core\texttrademark~i7 processor code named \textit{Nehalem} provides a feature named Turbo Boost which opportunistically varies the frequencies of the processor's cores.  The frequency of a core is determined by core temperature, the number of active cores, the estimated power consumption, the estimated current consumption, and operating system frequency scaling requests. For a chip multi-processor(CMP) that has a small number of physical cores and a small set of performance states, deciding the Turbo Boost frequency to use on a given core might not be difficult. However, we do not know the complexity of this decision making process in the context of a large number of cores, scaling to the 100s, as predicted by researchers in the field.
\section{Introduction}
\label{sec:intro}
In recent times, we have seen the introduction of  processors with multiple cores on chip, ranging from commodity 2, 4, and 8--core processors, the Polaris 80-core research processor from Intel, and Tilera's 64-core and recent 100-core offerings. Given these trends and the progression of Moore's Law, researchers are envisioning processors with 100s if not 1000s of cores on a single chip \cite{1531805, hill2008asl}. 

The architecture of such massively multicore chips poses new problems for scalability, both on hardware and software fronts, including performance, power consumption, memory and cache bottlenecks, software scalability to multiple threads, etc. Software written to take advantage of multiple cores, that is software that is multi-threaded is becoming increasingly prevalent. Various threading libraries that hide the complexities of multi-threading are being released - Cilk, Intel TBB, etc. However, all available software need not scale to use all the available cores on a multicore processor; legacy single threaded software might still be in use as it is today.

Given the above usage scenario, we can expect that some of the cores on a CMP will be idle and such cores can be transitioned to low power states (C states) and the frequency of busy cores can be boosted to provide improved performance. This behaviour forms the basis of the Turbo Boost feature present in the Intel\textregistered ~Core\texttrademark ~i7 processors (codenamed Nehalem)\cite{nehalemWhitePaper}. Turbo Boost is made possible by a processor feature named power gating. Traditionally, an idle processor core consumes little or no active power (which is due to transistor switching activity) while still dissipating static power due to its leakage current, even when it is operating at its lowest frequency. Power gating aims to cut the leakage current as  well, thereby further reducing the power consumption of the idle core. The extra power headroom available can be diverted to the active cores to increase their voltage and frequency without violating the power, voltage, and thermal envelopes. Turbo Boost is one way of providing performance boost to applications in a massively multicore processor setup. In this paper, we make a case that the current Turbo Boost mechanism is not sufficiently scalable and present a theoretical formulation of the problem involving optimal frequency assignment. 

In the remainder of the section, we present how Operating System Power Management (OSPM), Dynamic Voltage and Frequency Switching (DVFS), and Turbo Boost interact. In \secref{sec:deconstruct}, we present the results and analysis of our experiments intended to understand the behaviour of Turbo Boost. We present the optimal frequency assignment problem formulation in \secref{sec:prob-formulation} and related work in \secref{sec:related}. In the rest of the paper, we will ocassionally use the term Turbo instead of the full name, Turbo Boost.

\subsection{OSPM, DVFS \& Turbo Boost}
\label{sec:ospm}
Platform BIOS exports a P-State table that contains the performance state information -- a tuple containing the voltage and frequency identifiers -- that need to be written to hardware registers to change operating frequency of the processor core. The first entry in the table is referred to as $P_0$ and the last as $P_n$; $P_0$ corresponds  to the highest frequency and $P_n$ to the lowest frequency that the processor core can operate. \tabref{tab:withTurbo} and \tabref{tab:withoutTurbo} captures the frequencies reported by the BIOS with and without Turbo Boost respectively. These values were obtained from the \texttt{sysfs} entries exported by Linux, running on a quad-core Core\texttrademark~ i7 965 based machine. The Core i7 processor cores can operate at frequencies from 1.6 GHz to 3.4 GHz in steps of 133.33 MHz. Of the operating frequencies supported, one or more are marked as Turbo frequencies and the highest non-Turbo frequency is termed the \emph{maximum guaranteed frequency} ($P_{max}$). On the Core i7, 3.3 and 3.4 GHz are configured as Turbo frequencies and 3.2 is $P_{max}$. The 3.3 and 3.4 GHz frequencies are not reported directly to the OS. Instead, an operating point at 3193000 kHz is reported. This frequency is not supported by the processor and is the Turbo Boost indicator.

Under high load conditions, OSPM requests a frequency boost by writing the voltage and frequency identifiers into the appropriate CPU registers for the particular core. If the frequency requested is a Turbo Boost frequency, then firmware within the processor decides whether the operating frequency can be supported without violating the power and thermal constraints (listed earler), and if so, boosts the frequency to Turbo; Turbo Boost is an opportunistic feature.  If the processor cannot provide the Turbo frequency, it operates at the \emph{maximum guaranteed frequency}. Turbo Boost, therefore, is NOT under the control of OSPM and would appear quite non-deterministic from OSMP's perspective. 
\begin{table}[htbp]
\centering
\caption{Frequencies with Turbo}
\begin{tabular}{|c|l|}
  \hline
	P-State & Frequency (in kHz) \\
	\hline
	$P_0$ & 3193000\\
	\hline
	$P_1$ & 3192000\\
	\hline
	$\vdots$ & $\vdots$\\
	\hline
	$P_n$ & 1596000\\
	\hline
\end{tabular}
\label{tab:withTurbo}
\end{table}	
\begin{table}[htbp]
\centering
\caption{Frequencies \emph{without} Turbo}
\begin{tabular}{|c|l|}
  \hline
	P-State & Frequency (in kHz) \\
	\hline
	$P_0$ & 3192000\\
	\hline
	$\vdots$ & $\vdots$\\
	\hline
	$P_n$ & 1596000\\
	\hline
\end{tabular}
\label{tab:withoutTurbo}
\end{table}

\section{Deconstructing Turbo Boost}
\label{sec:deconstruct}
The Turbo Boost algorithm is proprietary and we do not have information about its internal workings. Therefore, we run a set of experiments and gather data, which we analyse to obtain information about the behaviour of Turbo. We run the BLAST ~\cite{citeulike:100088} benchmark  -- a suite of multithreaded applications which show varying rates of CPU utilization, and consequently could be expected to provide sufficient scope for Turbo Boost to engage. We measured CPU utilization using \texttt{mpstat}, and measured frequency using a frequency measurement tool developed in-house. Frequency calculation was done implementing the algorithm provided in \cite{turboWhitePaper} (also summarized below).
\begin{enumerate}
	\item The base operating ratio is obtained by reading the \texttt{PLATFORM\_INFO} Model Specific Register (MSR). The base operating ratio is multiplied by the bus clock frequency (133.33 MHz) to obtain the base operating frequency.
	\item The \texttt{Fixed Architectural Performance Monitor} counters are enabled. \texttt{Fixed Counter 1} counts the number of core cycles while the core is not in a halted state (\texttt{CPU\_CLK\_UNHALTED.CORE}). \texttt{Fixed Counter 2} counts the number of reference cycles when the core is not in a halted state (\texttt{CPU\_CLK\_UNHALTED.REF}).
	\item The two counters are read at regular intervals and the number of unhalted core cycles and unhalted reference cycles that have expired since the last iteration are obtained. Frequency is calculated as \\ $F_{current} = $ \texttt{Base Operating Frequency} $\times$ ( \texttt{Unhalted Core cycles} / \texttt{Unhalted Reference Cycles}). The frequency calculation is repeated for each core.
\end{enumerate}
For simplicity of analysis, we disabled Simultaneous Multi Threading (SMT) on the processor, therefore each physical core supported only one thread context. We run the benchmarks with the \texttt{ondemand} and the \texttt{userspace} frequency governors of Linux. The frequency governor, and the operating frequency in the case of \texttt{userspace} are set prior to starting the benchmark. 

The \texttt{ondemand} governor takes system dynamics into consideration and varies the processor operating frequency. Consequently, it is the most power-performance efficient policy, but its dynamism makes for difficult analysis when the effects of Turbo Boost are also included. In order to isolate the effects of Turbo Boost, we use the \texttt{userspace} governor and set the frequency of all cores to $P_{max}$. With this setup, the OSPM does not initiate frequency changes despite changes in CPU utilization, and all transitions in and out of the Turbo Boost frequncies are the effects of the Turbo Boost algorithm, without OSPM requests. 
\begin{figure*}
\centering
\includegraphics[scale=0.35]{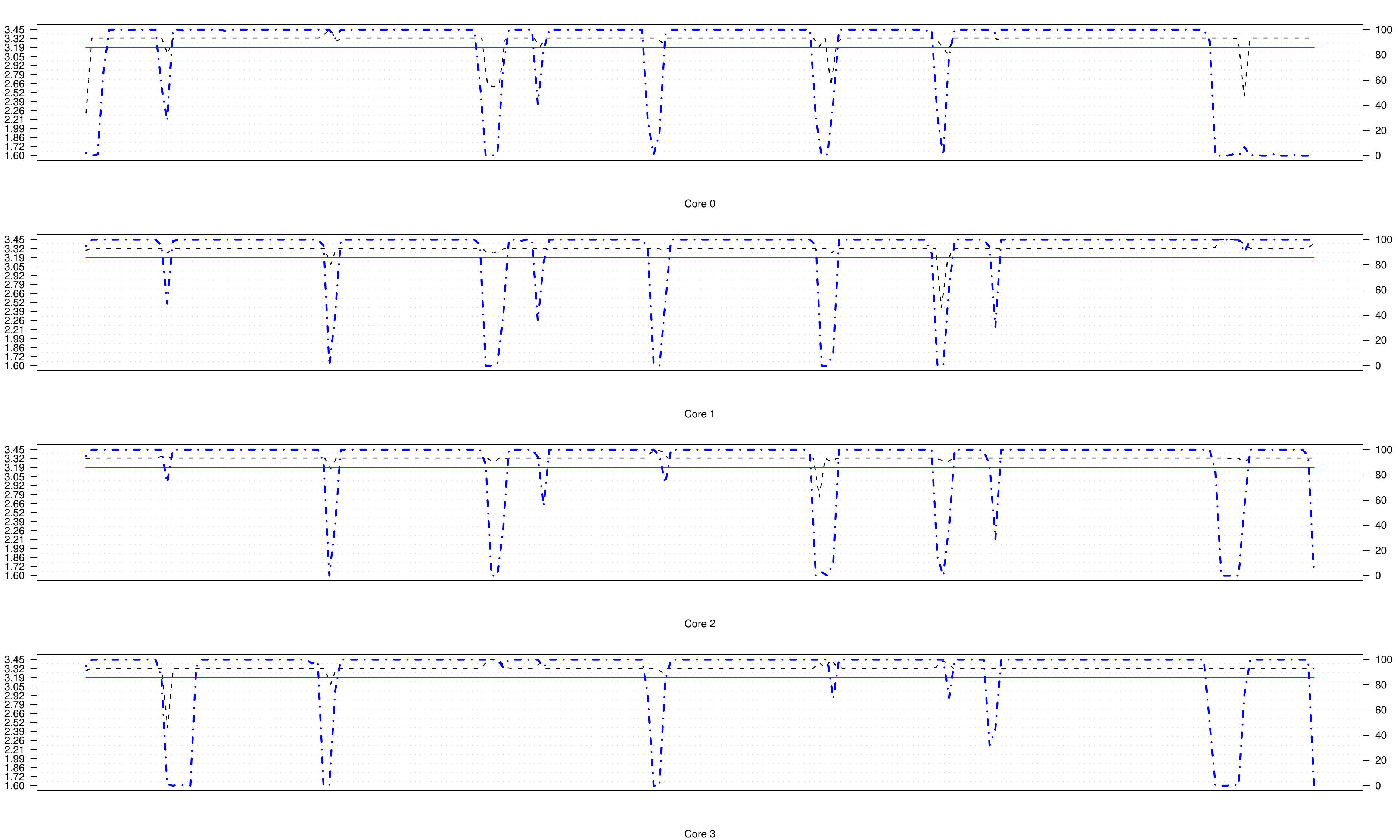}
\caption{tblastx -- \texttt{ondemand} governor \& default BIOS settings}
\label{fig:tblastx_defaults_ondemand}
\end{figure*}
\begin{figure*}
\centering
\includegraphics[scale=0.35]{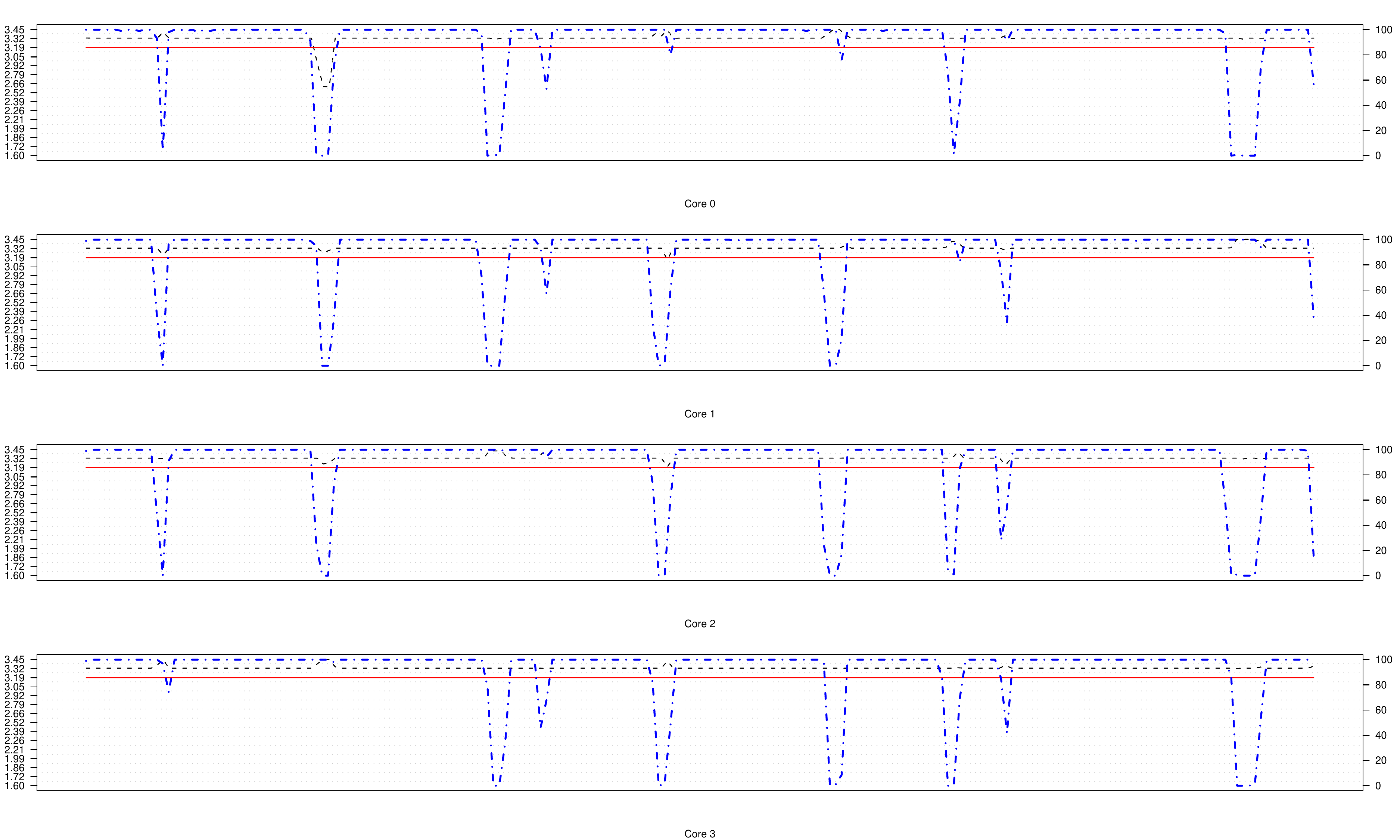}
\caption{tblastx -- \texttt{userspace} governor \& default BIOS settings}
\label{fig:tblastx_defaults_userspace}
\end{figure*}

The graphs in figures \ref{fig:tblastx_defaults_ondemand} and \ref{fig:tblastx_defaults_userspace} show the variation in frequency and utilization during one run of the \texttt{tblastx} benchmark from the BLAST suite with the \texttt{ondemand} and the \texttt{userspace} frequency governors respectively. In all the plots, the dashed line captures variation in frequency, whereas the the dot-dash line captures utilization. The solid horizontal line captures the maximum non Turbo Boost frequency (also refered to earlier as maximum guaranteed frequency or $P_{max}$).

Interestingly, we see very little variation in the frequency of the cores despite substantial variations in the CPU utilization. The \texttt{ondemand} governor does not appear to be particularly aggressive in pushing down frequencies of cores that are under low utilization. We do see however, that with all four cores active; the lower Turbo Boost frequency is reached and the higher Turbo Boost frequency is reached only when a single core is active. The \texttt{userspace} governor expectedly shows much lesser variation in core frequency. Except for one instance (for Core 0), the operating frequency of none of the cores goes below 3.2 GHz which we set prior to starting the benchmark. Even with the userspace governor, where all cores are operating at 3.2GHz frequency, we can observe that cores transition into the two turbo frequencies. 

\subsection{\textgreater 2 levels of Turbo}
\label{sec:manylevels}
\begin{figure*}
\centering
\includegraphics[scale=0.35]{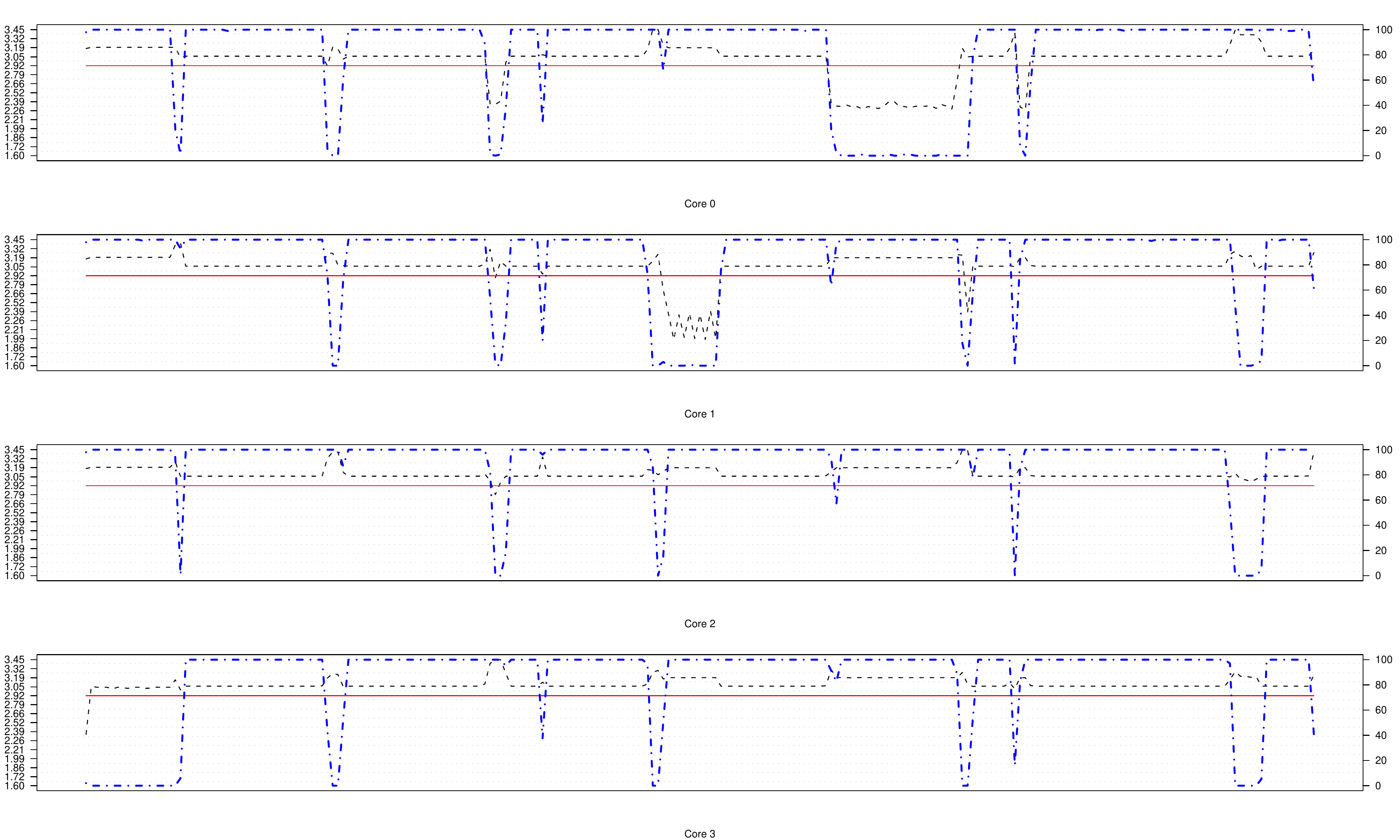}
\caption{tblastx -- \texttt{ondemand} governor \& modified BIOS settings}
\label{fig:tblastx_ondemand}
\end{figure*}
By default, the processor cores are configured to operate with a maximum guaranteed frequency of 3.2 GHz. However, these levels can be changed in the BIOS and for our second set of measurements, we set the maximum guaranteed frequency at 2.96 GHz, and all frequencies above it (that is four operating frequencies from 3.05 through 3.4 GHz) are considered as Turbo. We believe that understanding the behaviour of Turbo in such a  setup is important, because, in a massively multi-core CMP, it is likely to have cores that run at low frequency (by default), but also have multiple levels (more than just two) of Turbo Boost. With multiple applications executing on the massively multi-core processor, and each processor core going through various levels of utilization, different cores could be boosted to different Turbo levels as time progresses, provided that the power and thermal constraints (mentioned earlier) are not violated.

 \figref{fig:tblastx_ondemand} captures the variation of core frequency and CPU utilization.  Interestingly, the \texttt{ondemand} governor appears to become a lot smarter and more aggressive in reducing the frequency of idle cores. 
With the custom BIOS settings, we can expect that applications incur performance loss because the maximum guaranteed frequency is lower compared to the default BIOS settings and as expected, benchmarks suffered a performance loss, varying from 5.4\% reduction for \texttt{blastn} to 13.1\% reduction for \texttt{megablast}.
\begin{table}
\centering
\begin{tabular}{|c|c|}
\hline
Benchmark & \% Reduction in Performance\\
\hline
blastn & 5.4 \\
\hline
blastp & 9.6\\
\hline
blastx & 12.2\\
\hline
tblastn & 5.8\\
\hline
tblastx & 13.0\\
\hline
megablast & 13.1\\
\hline
\end{tabular}
\caption {Reduction in performance with custom BIOS settings}
\label{tab:perfLoss}
\end{table}

\subsection{Analysis \& Observations}
\label{sec:analysis}
From the figures presented, we can observe that when three cores are active, with the default BIOS settings, all the cores can simultaneously operate at 3.3 GHz. But with the modified BIOS settings, all the cores can operate simultaneously at a maximum frequency of 3.2 GHz. Consequently, we can infer that the Turbo Boost algorithm is governed by the  limits set in the BIOS. Strictly adhering to these limits could be acceptable and necessary from a reliability and repeatability standpoint.

However, such a setting could result in the processor operating at a lower frequency than what is possible without violating the constraints. From the example just described, we can see that with the modified BIOS settings, when 3 cores are active, the cores operate at 3.2 GHz though they could safely operate at 3.3 GHz as shown by our first set of experiments.  The hard limits result in sub-optimal frequency assignment and to unrealized performance. In a massively  multicore setup, the unrealized performance that accrues across all processor cores could be substantial. Nor is it feasible to manually set limits on such a processor. 

Further, we see instances where, despite the CPU utilization falling, Turbo Boost increases the frequency of cores. This behaviour suggests that DVFS requests play an important role, and if Turbo can be supported, the processor will do so, despite sub-optimality -- higher power consumption with no performance gain.  Therefore a scalable, dynamic, and efficient mechanism of frequency assignment is required. 

\section{Problem Formulation}
\label{sec:prob-formulation}

The best performance can be obtained when all active cores are operating at Turbo frequencies. However, this frequency assignment is impractical due to power, temperature, and voltage supply constraints within which the processor must operate. Let us consider a multicore processor with $n$ homogeneous cores - cores that have a common instruction set architecture (ISA), the same manufacturing process, and the same power-performance profiles. At any given point of time, $n_s$ cores  could be in a sleep state (idle), while the other $n_a$ cores are in active state. 

The processor supports a maximum guaranteed frequency (denoted  $f_0$) and some Turbo Boost frequencies $f_1,f_2,\ldots f_m$. At higher operating frequencies, a processor core consumes higher power which also results in higher processor temperature. In the following sections that discuss the problem formulation, we abandon the ACPI style nomenclature of frequencies of the form $P_i$ used earlier. 

\begin{equation}
\centering
P \alpha V^2f
\label{power}
\end{equation}

Every processor has a factory-specified maximum power limit $Pow_{max}$ and a critical shutdown temperature $T_{crit}$. It is well known that power consumption of a processor core is proportional to the supply voltage and frequency(Equation\ref{power}). The current processor temperature is represented as $\tau$ (below $T_{crit}$) and each core consumes power $p_0,p_1\ldots\ p_m$ for the corresponding frequencies. Two decisions need to be taken: for the OS the decision is to decide whether to activate one or more of the idle cores or to request Turbo Boost on the currently active cores; for the processor the decision is to identify the Turbo Boost frequency (if requested by OSPM) or ignore the OSPM request.

We assume that at all times, the processor keeps a count of the current number of active cores, and a list of estimated power consumption. For example consider a processor with 4 homogeneous cores with two cores active and two cores in sleep state. The Turbo Boost algorithm could have a list of estimated power consumption of the active cores as shown in \tabref{tab:ex-power-table}. If we stipulate that all the cores \textit{must} operate at the same frequency at any given point in time, this problem can be trivially solved by performing a table lookup. For instance, if  two cores are active and the maximum allowed power consumption is 132W, then we can easily lookup the table and obtain the Turbo frequency as 2.5 GHz.
\begin{table}[htd]
\centering
\begin{tabular}{ccc}
\hline
Active & \multicolumn{2}{c}{Turbo Frequency} \\
Cores & (2.3GHz) & (2.5GHz) \\
\hline
1 & 129 & 131\\
2 & 130 & 132\\
3 & 136 & 138\\
4 & 140 & 142\\
\hline
\end{tabular}
\caption{Power estimates of active cores}
\label{tab:ex-power-table}
\end{table}

However, for processors where cores can operate at different frequencies the problem is considerably difficult. We must chose an assignment of different frequency values to different cores subject to the power consumption constraint, such that the processing power is maximized. We can model this optimization problem as the following Integer Linear Program (ILP). 

Let there be binary variables $x_{ij}$ which take the value $1$ when core $i$ is assigned frequency $f_j$. Then, the power contribution of the core is $x_{ij}\times p_j$ and performance contribution is $x_{ij}\times f_j$ and we can formulate the following maximization problem. 
\begin{align*}
\displaystyle \text{Maximize} & \sum_{i\in n}\sum_{j\in m}{x_{ij}f_j}\\
\text{ such that } & \sum_{i\in n}\sum_{j\in m}{x_{ij}p_j}\leq Pow_{max}\tag{1a}\label{q:1a}\\
\forall i\in n, & \sum_{j\in m} x_{ij}\leq 1\tag{1b}\label{q:1b}\\
\end{align*}
This is well known generalization of the Knapsack Problem. Additionally, if the items are divided into classes and we are allowed to chose at most one item from each class. This modified problem is a case of the multiple choice knapsack problem (MCKP). When the objective function is a constant, that is, we are looking to achieve a particular performance value (as in our problem context), then the problem reduces to the subset-sum problem. Since both the MCKP and subset-sum problems are NP-Complete, we have little hope of finding an algorithm to calculate the exact solution in polynomial time. 

However, we can devise an approximation algorithm that can run in a short time interval and provide the frequency assignment. Naturally, the frequency assignment will be sub-optimal from a performance point of view, but we can expect the algorithm to outperform the existing algorithm and never violate power and thermal constraints. 

Additional factors that could influence Turbo Boost are temperature; however, we did not observe a strong correlation between temperature and Turbo Boost entry/exit therefore we did not incorporate temperature as a constraint in our problem formulation. Further, the current formulation does not incorporate task assignment or task properties. The problem we have presented however, is not restricted to frequency assignment across a large number of cores; it is a problem of frequency assignment to obtain the maximum performance for all applications under the various constraints. To that effect, we can augment the frequency assignment such that cores are selectively Turbo Boosted depending on the application that is running on the core, where, cores running CPU bound applications (which see a greater benefit from increased frequency) are boosted,  whereas cores running memory bound applications (which do not see much benefit due to to frequency boosts)~\cite{1531804} are not boosted despite requests from OSPM to go into highest frequency. Such an approach would require information to be obtained from the hardware performance counters on each core. 

\section{Related Work}
\label{sec:related}
DVFS has been extensively used in many studies on power management. DVFS has been used in a variety of power management mechanisms, and the work by Govil et al ~\cite{215546} being one of the earliest, where they compare algorithms for setting the CPU frequency. Ishihara et. al ~\cite{280894}, investigate optimal voltage assignment (formulated as an ILP). However, their approach requires information like the number of cycles of each task, the average switched capacitance per task, etc. These metrics would have to be gathered off line, consequently, it would not be possible to use the approach when a completely new workload is presented. 

Dhiman et al \cite{1233656} propose a machine learning based approach to vary dynamic power management policies and adapts to to varying system workloads. Their approach is a system level approach and includes peripheral devices in the power management scheme, specifically focusing on hard disk and wireless LAN device power management; CPU power management is not taken into consideration.

In ~\cite{1289845}, Murali et al study optimal, temperature-aware frequency assignment for multi-processor System on Chips (MPSoCs) in which they model and account for heat flow between parts of the MPSoC and use convex optimization for obtaining the optimal frequency assignment. However, we focus particularly on the Turbo Boost feature. Further, their work incorporates lot of details from the physics and VLSI design process into the formulation. This is justifiable from their goal which is to aid chip designers with an estimate of the power consumption characteristics. In contrast, our model tries to do the optimization at the firmware or OS level which is after processor fabrication and manufacturing. At this stage, the theoretical estimates of power and capacitance are not that useful because for a particular design or architecture the values may differ from gross estimates and the vendor provided thermal ratings are a closer approximation. 

The work by Isci et al \cite{1194850} comes closest to our work. Isci et al investigate per core and chip wide DVFS policies to achieve maximum performance within a given power budget. They evaluate three local polices: \textit{Priority}, \textit{PullHiPushLo}, and \textit{MaxBIPS} and a global power policy. However, their evaluation considers a processor that has a small number of operating states; also, they do not use an ILP formulation. We are interested in understanding the complexity of scaling a frequency assignment to a very large number of cores and providing optimal frequency and performance without violating the power and thermal constraints, which inherently leads to the ILP formulation we presented earlier. 
\nocite{*}
\bibliographystyle{abbrv}
\bibliography{TurboScaling}
\end{document}